\documentclass[conference, a4paper, 9pt]{IEEEtran}

\IEEEoverridecommandlockouts
\usepackage{cite}
\usepackage{amsmath,amssymb,amsfonts}
\usepackage{algorithmic}
\usepackage{graphicx}
\usepackage{textcomp}
\usepackage{xcolor}
\def\BibTeX{{\rm B\kern-.05em{\sc i\kern-.025em b}\kern-.08em
    T\kern-.1667em\lower.7ex\hbox{E}\kern-.125emX}}
\usepackage{enumitem}
\usepackage{booktabs}
\usepackage{multirow}
\usepackage[super]{nth}
\usepackage{tikz}

\usetikzlibrary{tikzmark}
\tikzset{circledColor/.style={circle,draw,inner sep=0.1em,line width=0.04em}}
\usepackage{setspace}

\usepackage{fancyhdr}
\fancypagestyle{firstpage}{
  \fancyhf{}
  \chead{Accepted at the IEEE Biomedical Circuits and Systems Conference (BioCAS), October 16th - 18th, 2025 in Abu Dhabi, UAE.}
  \cfoot{\thepage}
}

\begin{document}

\title{SpikeVox: Towards Energy-Efficient Speech Therapy Framework with Spike-driven Generative Language Models
\vspace{-0.4cm}}


\author{
\IEEEauthorblockN{Rachmad Vidya Wicaksana Putra$^{1}$, Aadithyan Rajesh Nair$^{2}$, Muhammad Shafique$^{1}$}
\IEEEauthorblockA{$^{1}$\textit{eBrain Lab, New York University (NYU) Abu Dhabi, Abu Dhabi, UAE} \\
$^{2}$\textit{Pune University, Pune, India}}
\{rachmad.putra, muhammad.shafique\}@nyu.edu, nair.aadithyan999@gmail.com
\vspace{-0.6cm}
}

\maketitle
\pagestyle{plain}
\thispagestyle{firstpage}

\begin{spacing}{0.96}
\begin{abstract}
Speech disorders can significantly affect the patients' capability to communicate, learn, and socialize.   
However, existing speech therapy solutions (e.g., therapist or tools) are still limited and costly, hence such solutions remain inadequate for serving millions of patients worldwide. 
To address this, state-of-the-art methods employ neural network (NN) algorithms to help accurately detecting speech disorders.
However, these methods do not provide therapy recommendation as feedback, hence providing partial solution for patients.
Moreover, these methods incur high energy consumption due to their complex and resource-intensive NN processing, hence hindering their deployments on low-power/energy platforms (e.g., smartphones).   
Toward this, we propose \textit{SpikeVox}, a novel framework for enabling energy-efficient speech therapy solutions through spike-driven generative language model.  
Specifically, SpikeVox employs a speech recognition module to perform highly accurate speech-to-text conversion; leverages a spike-driven generative language model to efficiently perform pattern analysis for speech disorder detection and generates suitable exercises for therapy; provides guidance on correct pronunciation as feedback; as well as utilizes the REST API to enable seamless interaction for users.  
Experimental results demonstrate that SpikeVox achieves 88\% confidence level on average in speech disorder recognition, while providing a complete feedback for therapy exercises. 
Therefore, SpikeVox provides a comprehensive framework for energy-efficient speech therapy solutions, and potentially addresses the significant global speech therapy access gap. 
\end{abstract}


\section{Introduction}
\label{Sec_Intro}

Speech disorders make the persons (patients) having difficulty in producing proper sounds when speaking.
Such disorders affect nearly 1 in 12 US children (ages 3-17 years old) with nearly half of them have not received intervention services, and make more than 3 millions US people stutter~\cite{ASHA}\cite{NIDCD}.
Such disorders significantly affect the patients' capability to communicate, learn, and socialize, which often lead to difficulties in adapting to their personal and professional/career life~\cite{beaminghealth}.
To address this, speech therapy is required. 
It involves assessing, diagnosing, and treating speech disorders (e.g., articulation, fluency, resonance, and expressive disorders)~\cite{AHPC}.

Traditional speech therapy typically requires one-on-one sessions with speech-language pathologists (SLPs), who will guide patients through customized exercises designed to improve specific speech disorders~\cite{beaminghealth}\cite{Greatspeech}.
However, getting a session with SLP is very limited and costly (e.g., \$100 to \$250 per-hour)~\cite{beaminghealth}. 
Moreover, it often requires multiple sessions to complete the treatments. 
According to the World Health Organization (WHO), $\sim$1 billion people worldwide require speech therapy services, but only about 10\% have access to qualified providers. 
These data show that the traditional speech therapy remains inadequate for addressing patients worldwide.

\textbf{Targeted Research Problem:} 
\textit{How can we develop an automated speech therapy solution that provides highly accurate speech disorder detection and suggests suitable treatments?}
An efficient solution to this problem may enable a low-cost speech therapy solution that is accessible for patients worldwide. 
    
\subsection{State-of-the-art in Speech Therapy and Their Limitations}
\label{Sec_Intro_SOTA}

Currently, state-of-the-art methods employ neural network (NN) algorithms to accurately detect speech disorders, thereby helping human SLPs to identify different types of speech disorders~\cite{Ref_Deka_ReviewSpeechTherapy_SLH25, Ref_Bayerl_Dysfluencies_arXiv22, Ref_Mulfari_DL4Telerehab_CMB22, Ref_Shih_Dysarthria_Healthcare22}.
For instance, recent works proposed stuttering detection techniques by leveraging the wav2vec 2.0 library~\cite{Ref_Bayerl_Dysfluencies_arXiv22}\cite{Ref_Baevski_wav2vec_NeurIPS20}, keyword recognition using a trained deep learning model~\cite{Ref_Mulfari_DL4Telerehab_CMB22}, and dysarthria detection using convolutional neural networks (CNNs)~\cite{Ref_Shih_Dysarthria_Healthcare22}. 
However, these methods provide partial solution for patients, as they only focus on the speech disorder detection aspect and do not provide recommended treatments as feedback. 
Therefore, they still involve SLPs in the loop to guide the patients with necessary treatments, which limit the accessibility of the solution worldwide. 
Moreover, these methods incur high energy consumption due to their complex and resource-intensive NN processing, hence hindering their deployments for low-power/energy platforms (e.g., smartphones, embedded platforms, or wearable devices), which are especially important if patients require offline processing due to better efficiency and better privacy.

\vspace{-0.1cm}
\subsection{Associated Research Challenges}
\label{Sec_Intro_Challenges}
\vspace{-0.1cm}

The above-discussed limitations expose several characteristics that are expected from speech therapy solutions, as follows.
The solution should (1) detect speech disorders and categorize them with high accuracy, (2) provide feedback of recommended treatments/exercises based on the detected speech patterns, and (3) process NN algorithms efficiently to enable its adoption in low-power/energy systems, such as smartphones, embedded platforms, or wearable devices.

\vspace{-0.1cm}
\subsection{Our Novel Contributions}
\label{Sec_Intro_Novelty}
\vspace{-0.1cm}

\begin{figure*}[t]
\centering
\includegraphics[width=0.82\linewidth]{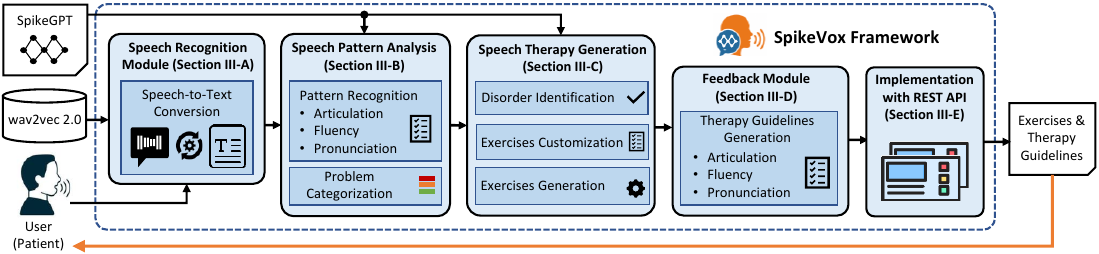}
\vspace{-0.3cm}
\caption{Overview of our SpikeVox framework, with its key design components highlighted in blue.} 
\label{Fig_SpikeVox}
\vspace{-0.5cm}
\end{figure*}

To address the challenges, we propose \textit{SpikeVox}, \textit{a novel framework for enabling energy-efficient speech therapy solutions by leveraging spike-driven generative language model.}
This paper is also the first work that provides a complete speech disorder detection, analysis, and feedback for speech therapy in a single spiking-based framework. 
To achieve this, SpikeVox employs the following key design steps (an overview shown in Fig.~\ref{Fig_SpikeVox}). 
\begin{itemize} [leftmargin=*]
    \item \textbf{Speech Recognition Module (Section~\ref{Sec_SpikeVox_SpeechRecog}):}
    It aims to capture the speech sound from the patient, and then perform speech-to-text conversion while preserving pronunciation information.
    \item \textbf{Speech Pattern Analysis (Section~\ref{Sec_SpikeVox_Analysis}):}
    It identifies errors in speech patterns using binary spike activations, then classifies them into the correct disorder categories.  
    \item \textbf{Speech Therapy Generation (Section~\ref{Sec_SpikeVox_Therapy}):}
    It generates customized exercises based on the detected disorder categories, hence providing effective therapy for the observed disorder. 
    \item \textbf{Feedback Module (Section~\ref{Sec_SpikeVox_Feedback}):}
    It provides guidance on the correct pronunciation based on the generated exercises, and thereby enabling the user to perform treatments without personal assistant.  
    \item \textbf{Implementation using the REST API (Section~\ref{Sec_SpikeVox_API}):}
    It aims to enable seamless interaction between SpikeVox system and the user through standard HTTP requests. 
\end{itemize}

\textbf{Key Results:}
In evaluation, we realize our SpikeVox framework using Python-based implementation, and then run it on the Apple M4 10-core CPU with 16GB memory.  
Experimental results show that, SpikeVox obtains high confidence level (88\% on average) in speech disorder recognition, while providing therapy exercises as feedback.

\section{Related Work}
\label{Sec_RelatedWork}

\textbf{Spiking Neural Networks (SNNs):}
An SNN model comprises several components, i.e., spiking neurons, network architecture, neural coding, and learning rule~\cite{Ref_Maass_SNN_NeuNet97, Ref_Mozafari_SpykeTorch_FNINS19, Ref_Putra_FSpiNN_TCAD20, Ref_Putra_QSpiNN_IJCNN21, Ref_Putra_SparkXD_DAC21, Ref_Putra_SpikeDyn_DAC21}.
Recently, SNNs have emerged as the alternate low-power/energy NN algorithms due to their sparse spike-driven operations~\cite{Ref_Shafique_EdgeAI_ICCAD21, Ref_Putra_QSViT_arXiv25, Ref_Putra_SNNonCNP_arXiv25, Ref_Putra_SpikeNAS_TAI25} and hardware advancements~\cite{Ref_Akopyan_TrueNorth_TCAD15, Ref_Davies_Loihi_MM18, Ref_Frenkel_ODIN_TBCAS19, Ref_SynSense_DYNAP, Ref_BrainChip_Akida}. 
Therefore, \textit{in this work, we leverage SNNs to perform speech disorder recognition and generate suitable exercises for therapy.}

\textbf{Spike-driven Generative Language Models:}
In this work, we use the spike-driven generative model called \textit{SpikeGPT}~\cite{Ref_Chu_SpikeGPT_TMLR24}, and employ its pre-trained model based on the OpenWebText2 dataset~\cite{Ref_Gao_Pile_arXiv20}. 
This model considers the Leaky Integrate-and-Fire (LIF) as the spiking neuron model, since it is commonly adopted in the SNN community due to its low computational complexity~\cite{Ref_Izhikevich_CompareModels_TNN04, Ref_Putra_ReSpawn_ICCAD21, Ref_Putra_lpSpikeCon_IJCNN22, Ref_Putra_SoftSNN_DAC22, Ref_Putra_EnforceSNN_FNINS22, Ref_Putra_TopSpark_IROS23, Ref_Rathi_SNNsurvey_CSUR23}.
For neural coding and learning rule, it employs rate coding~\cite{Ref_Diehl_STDPmnist_FNCOM15} and surrogate gradient learning~\cite{Ref_Neftci_SurrogateSNNs_IEEEMSP19}, respectively.
For network architecture, SpikeGPT uses Spiking Receptance Weighted Key Value (Spiking RWKV) and Spiking Receptance Feed-Forward Networks (Spiking RFFN) modules.

\section{The SpikeVox Framework}
\label{Sec_SpikeVox}

Fig.~\ref{Fig_SpikeVox} shows our SpikeVox framework with its novel key components, which are further discussed in the following subsections.

\subsection{Speech Recognition Module}
\label{Sec_SpikeVox_SpeechRecog}

This step captures the speech sound from the patient, and then perform speech-to-text conversion while preserving pronunciation information.
Unlike conventional speech-to-text techniques which only focus on semantic accuracy, SpikeVox captures both the transcription and confidence scores for individual phoneme.
This speech recognition module leverages the wav2vec 2.0 library~\cite{Ref_Baevski_wav2vec_NeurIPS20} for phoneme-level analysis. 
By examining the softmax output of the model, we identify potential pronunciation issues where confidence is low. 
This information is then passed to the pattern analysis step.

\subsection{Speech Pattern Analysis}
\label{Sec_SpikeVox_Analysis}

This step identifies disorders in speech patterns using binary spike activations. 
Specifically, it leverages the pre-trained SpikeGPT model to analyze articulation, fluency, and pronunciation of the input sound, and then categorize the detected issues based on common speech therapy classification:
(1) \textit{R-sound issues (rhotacism)}, (2) \textit{S-sound issues (lisping)}, (3) \textit{Th-sound issues}, (4) \textit{L-sound issues}, (5) \textit{consonant cluster simplification}, and (6) \textit{vowel distortions}~\cite{Ref_Shriberg_DisorderClass_CLP10}.

For each category, the system assigns a confidence score, creating a comprehensive profile of the patient's speech patterns. This profile serves as the basis for generating personalized therapy exercises.
The confidence score $C_i$ for speech disorder category $i$ is defined as:
\begin{equation}
    C_i = \alpha \cdot \frac{\sum_{j \in \mathcal{P}_i} (1-p_j)}{\left|\mathcal{P}_i\right|} + \beta \cdot \frac{S_i}{S_{max}} + \gamma \cdot M_i
    \label{Eq_confidence}
\end{equation}
Here, $\mathcal{P}_i$ is the set of phonemes associated with category $i$; $p_j$ is the confidence score from the wav2vec 2.0 for phoneme $j$; $S_i$ is the spike density for neurons associated with category $i$; $S_{max}$ is the maximum possible spike density; $M_i$ is the pattern matching score derived from membrane potential of SpikeGPT model; while $\alpha$, $\beta$, and $\gamma$ are weighting factors with $\alpha + \beta + \gamma = 1$. 

The contribution of SpikeGPT to confidence scoring is realized through the spike density ($S_i$) and the pattern matching score ($M_i$). 
Spike density $S_i$ measures how frequently category-specific neurons activate when processing problematic phonemes, and it is defined as:
\begin{equation}
     S_i = \frac{\sum_{t=1}^{T} \sum_{n \in \mathcal{N}_i} s_{n,t}}{T \cdot |\mathcal{N}_i|}
\end{equation}
Here, $s_{n,t} \in \{0,1\}$ is the binary spike output of neuron $n$ at time step $t$; $\mathcal{N}_i$ is the set of neurons associated with category $i$; and $T$ is the total number of time steps.
Meanwhile, pattern matching score $M_i$ leverages the membrane potential patterns in SpikeGPT by comparing them to known disorder patterns, and it can be expressed as:
\begin{equation}
     M_i = 1 - \frac{1}{|\mathcal{N}_i|} \sum_{n \in \mathcal{N}_i} \mathrm{sim}(U_{n}, \hat{U}_{i,n})
\end{equation}
Here, $U_{n}$ is the membrane potential sequence for neuron $n$; $\hat{U}_{i,n}$ is the reference membrane potential pattern for disorder category $i$ and neuron $n$; and $\mathrm{sim}(\cdot)$ is a similarity function such as cosine similarity.

Once the confidence score is calculated, it will be mapped to a specific disorder based on the neuron activation patterns; see Table~\ref{tab_mapping_categories}. 

\begin{table}[h]
\vspace{-0.3cm}
\centering
\caption{Mapping speech disorder categories based on the neuron activation patterns of SpikeGPT.}
\label{tab_mapping_categories}
\vspace{-0.1cm}
\scriptsize
\begin{tabular}{|c|c|c|c|c|}
\hline
\textbf{Disorder} & \textbf{Primary} & \textbf{Typical} &
\textbf{Threshold} & \textbf{Weight} \\
\textbf{Category} & \textbf{Neurons} & \textbf{$S_i$} &
 & \textbf{($\alpha$, $\beta$, $\gamma$)} \\
\hline
\hline
R-sound issues & N$_{1-64}$ & 0.15-0.35 & 0.25 & (0.5, 0.3, 0.2) \\
S-sound issues & N$_{65-128}$ & 0.20-0.40 & 0.30 & (0.4, 0.4, 0.2) \\
Th-sound issues & N$_{129-192}$ & 0.25-0.45 & 0.35 & (0.4, 0.3, 0.3) \\
L-sound issues & N$_{193-256}$ & 0.15-0.30 & 0.20 & (0.5, 0.3, 0.2) \\
Consonant clusters & N$_{257-320}$ & 0.30-0.50 & 0.40 & (0.3, 0.5, 0.2) \\
Vowel distortions & N$_{321-384}$ & 0.20-0.35 & 0.25 & (0.4, 0.3, 0.3) \\
\hline
\end{tabular}
\vspace{-0.2cm}
\end{table}

\vspace{-0.1cm}
\subsection{Speech Therapy Generation}
\label{Sec_SpikeVox_Therapy}

This step leverages language generation features in the SpikeGPT to produces contextually appropriate practice sentences that focus on problematic phonemes and sound combinations. 
This step considers:
(1) severity of each identified issue; (2) phonetic context in which errors occur; (3) progression from simpler to more complex exercises; and (4) personalization based on patient history and progress.

The exercise generation process is formulated as an optimization problem over a set of candidate sentences. For a given speech disorder category $c$, difficulty level $d$, and patient history vector $\mathbf{h}$, we define the optimal exercise selection function $\mathcal{E}(c, d, \mathbf{h})$ as:
\begin{equation}
    \mathcal{E}(c, d, \mathbf{h}) = \underset{s \in \mathcal{S}}{\arg\max} \Big[ \omega_1 \cdot \mathcal{R}(s, c) + \omega_2 \cdot \mathcal{D}(s, d) + \omega_3 \cdot \mathcal{P}(s, \mathbf{h}) \Big]
\end{equation}
Here, $\mathcal{S}$ represents the set of potential exercise sentences; 
$\mathcal{R}(s, c)$ is the relevance function measuring how well sentence $s$ targets disorder category $c$; $\mathcal{D}(s, d)$ is the difficulty alignment function for difficulty level $d$; $\mathcal{P}(s, \mathbf{h})$ is the personalization function based on patient history $\mathbf{h}$; while $\omega_1, \omega_2$, and $\omega_3$ are weighting parameters where $\omega_1 + \omega_2 + \omega_3 = 1$.

The relevance function $\mathcal{R}(s, c)$ quantifies how effectively a sentence targets the specific speech disorder category:
\begin{equation}
    \mathcal{R}(s, c) = \lambda_c \cdot \frac{\sum_{i=1}^{|s|} \psi(s_i, \Phi_c) \cdot (1 + \eta \cdot \mathcal{Q}(s_i, s_{i-1}, s_{i+1}))}{|s|}
\end{equation}
Here, $\psi(s_i, \Phi_c)$ is an indicator function that equals 1 if phoneme $s_i$ belongs to the target phoneme set $\Phi_c$ for category $c$, and 0 otherwise; $\mathcal{Q}(s_i, s_{i-1}, s_{i+1})$ is a contextual difficulty factor that increases the score when the target phoneme appears in challenging phonetic contexts; $\lambda_c$ is a category-specific normalization constant; while $\eta$ is a context weighting parameter.

The difficulty alignment function $\mathcal{D}(s, d)$ ensures that generated exercises match the desired difficulty level, and can be expressed as:
\begin{equation}
    \mathcal{D}(s, d) = 1 - \Big|\frac{\mathcal{C}(s) - \mu_d}{\delta_d}\Big|
\end{equation}
Here, $\mathcal{C}(s)$ measures the complexity of sentence $s$; $\mu_d$ is the target complexity for difficulty level $d$; and $\delta_d$ controls the acceptable deviation from the target. The function $\mathcal{C}(s)$ is defined as:
\begin{equation}
    \mathcal{C}(s) = \alpha_1 \cdot |s| + \alpha_2 \cdot \frac{|\mathcal{V}_s|}{|s|} + \alpha_3 \cdot \mathcal{CC}(s) + \alpha_4 \cdot \mathcal{SR}(s)
\end{equation}
$|s|$ is the sentence length; $|\mathcal{V}_s|$ is the vocabulary size in the sentence; $\mathcal{CC}(s)$ is the consonant cluster density; $\mathcal{SR}(s)$ is the syllabic rhythm complexity; while $\alpha_1, \alpha_2, \alpha_3$, and $\alpha_4$ are weighting coefficients.

The personalization function $\mathcal{P}(s, \mathbf{h})$ adapts exercises based on the patient's history, and can be expressed as:
\begin{equation}
    \mathcal{P}(s, \mathbf{h}) = \gamma_1 \cdot \text{sim}(s, \mathbf{h}_\text{success}) \cdot (1 - \gamma_2 \cdot \text{sim}(s, \mathbf{h}_\text{failure}))
\end{equation}
Here, $\text{sim}(s, \mathbf{h}_\text{success})$ measures similarity between sentence $s$ and previously successful exercises; $\text{sim}(s, \mathbf{h}_\text{failure})$ measures similarity to previously failed exercises; and $\gamma_1, \gamma_2$ are weighting parameters.

To generate candidate sentences using SpikeGPT, we construct category-specific prompts $\mathcal{G}(c, d)$ that guide the language model, with template-based fallbacks ensuring clinical quality:
\begin{equation}
    \mathcal{G}(c, d) = \text{prefix}_c \oplus \text{modifier}_d \oplus \text{instruction}_c
\end{equation}
Here, $\text{prefix}_c$ is a category-specific prefix (e.g., "Create a sentence with many R sounds:"); $\text{modifier}_d$ adjusts for difficulty (e.g., "Make it short and simple" for easy); $\text{instruction}_c$ provides specific phonetic guidance; and $\oplus$ represents string concatenation.

The generation process involves sampling from SpikeGPT's output distribution with temperature $\tau_d$ that varies with difficulty level:
\begin{equation}
    s \sim \mathcal{M}(\mathcal{G}(c, d), \tau_d, \kappa)
\end{equation}
$\mathcal{M}$ represents the SpikeGPT model; $\tau_d$ is the sampling temperature for difficulty level $d$; and $\kappa$ is a parameter controlling output diversity.
However, given the domain gap between general language training and therapy-specific text, a quality filtering mechanism selects clinically appropriate alternatives when generated output quality is insufficient for therapeutic use.

\vspace{-0.2cm}
\subsection{Feedback Module}
\label{Sec_SpikeVox_Feedback}
\vspace{-0.1cm}

This step provides personalized guidance on the correct pronunciation based on the generated exercises, and thereby enabling the user to perform treatments without personal assistant.
It generates three types of feedback: (1) specific phoneme-level guidance for detected issues, (2) visual pronunciation guides showing tongue and lip positions, and (3) general practice recommendations to improve overall articulation.

The feedback generation process takes two primary inputs: speech analysis results ($\mathcal{A}$), and exercise performance results ($\mathcal{E}$) when available. 
The feedback function ($\mathcal{F}$) can be formally defined as:
\begin{equation}
    \mathcal{F}(\mathcal{A}, \mathcal{E}) = \{\mathcal{F}_s, \mathcal{F}_g, \mathcal{F}_v, \mathcal{F}_o, \mathcal{F}_e\}
\end{equation}
$\mathcal{F}_s$, $\mathcal{F}_g$, $\mathcal{F}_v$, $\mathcal{F}_o$, and $\mathcal{F}_e$ denote specific guidance, general tips, visual guides, overall assessment, and exercise-specific feedback, respectively.
To generate $\mathcal{F}_s$, we select from category-specific templates ($\mathcal{T}_c$) based on the detected issues ($\mathcal{I}$), stated as: 
\begin{equation}
    \mathcal{F}_s = \{(c, \text{select}(\mathcal{T}_c)) \mid c \in \mathcal{I}\}
\end{equation}
Here, $\text{select}(\mathcal{T}_c)$ is a selection function that chooses an appropriate guidance from the set of templates $\mathcal{T}_c$ for category $c$.
Visual guides $\mathcal{F}_v$ are generated based on the phonetic categories requiring attention:
\begin{equation}
    \mathcal{F}_v = \{(t_c, d_c, r_c) \mid c \in \mathcal{I}\}
\end{equation}
$t_c$ is the guide type (e.g., tongue position), $d_c$ is the description, and $r_c$ is the reference to the visual guide for category $c$.

SpikeVox integrates confidence-weighted guidance, hence its feedback is prioritized based on the severity of disorder and the confidence in system's detection, ensuring that patients receive the most effective guidance.
When exercise performance data $\mathcal{E}$ is available, the system calculates an accuracy score ($\mathcal{A}_c$) for each exercise category:
\begin{equation}
    \mathcal{A}_c = \mathcal{A}_{\text{base},c} + \lambda_c \cdot \mathcal{A}_{\text{adj},c}
\end{equation}
$\mathcal{A}_{\text{base},c}$ is a baseline accuracy for category $c$; $\mathcal{A}_{\text{adj},c}$ is an adjustment factor; and $\lambda_c$ is a category-specific weighting parameter. 
These accuracy scores are then mapped to qualitative assessments through a thresholding function:
\begin{equation}
    \mathcal{F}_e = \{\mathcal{A}_c, \text{assess}(\mathcal{A}_c), \mathcal{I}_{\text{areas}}, \mathcal{S}_{\text{strengths}}\}
\end{equation}
Here, $\text{assess}(\mathcal{A}_c)$ maps the accuracy score to a textual assessment, $\mathcal{I}_{\text{areas}}$ identifies improvement areas, and $\mathcal{S}_{\text{strengths}}$ highlights strengths in the patient's performance.
The overall assessment $\mathcal{F}_o$ is determined by the severity level $\sigma$ identified in the analysis:
\begin{equation}
    \mathcal{F}_o = 
    \begin{cases}
        \text{``Simple practice''}, & \text{if } \sigma = \text{``mild''} \\
        \text{``Focused practice''}, & \text{if } \sigma = \text{``moderate''} \\
        \text{``Intensive practice''}, & \text{if } \sigma = \text{``severe''}
    \end{cases}
    \label{Eq_severity}
\end{equation}
Here, we define that a sample is considered as ``mild'', ``moderate'', and ``severe'' if its number of issues (\#issues) meets: (1) \#issues$\leq$5, (2) 5$<$\#issues$\leq$10, and \#issues$>$10, respectively.

\vspace{-0.1cm}
\subsection{Implementation using the REST API}
\label{Sec_SpikeVox_API}

\begin{figure*}[t]
\centering
\includegraphics[width=\linewidth]{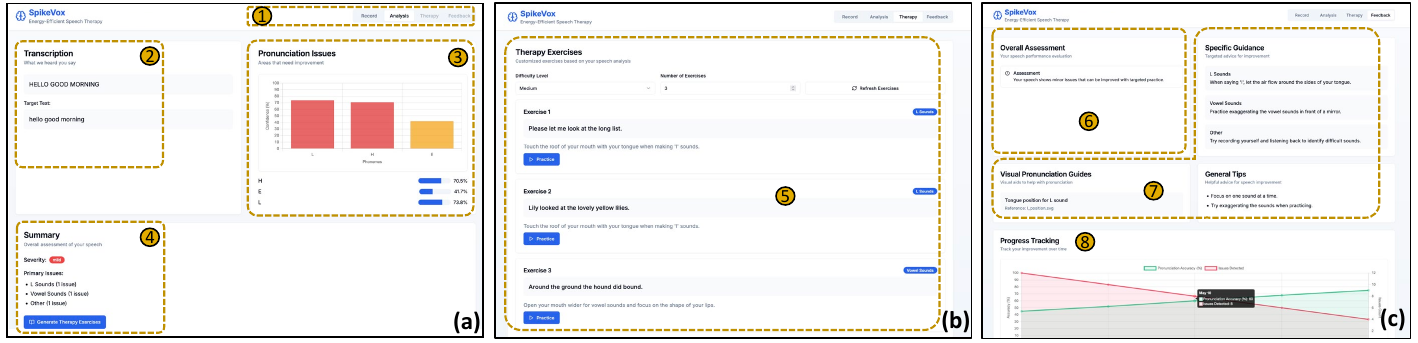}
\vspace{-0.6cm}
\caption{Display of the SpikeVox's dashboard for (a) analysis page; (b) therapy generation page; and (c) feedback page.} 
\label{Fig_App}
\vspace{-0.4cm}
\end{figure*}

To enable seamless interaction with the SpikeVox system, we integrate SpikeVox components using the REST Application Programming Interface (API) through standard HTTP requests. 
Moreover, this API also facilitates integration with various front-end applications, making the system adaptable to different use cases and platforms.
This REST API implementation has three main endpoints, as follows.
\begin{itemize}[leftmargin=*]
    \item \texttt{/api/speech-analyze}: 
    It processes audio input and returns a detailed analysis of speech patterns.
    \item \texttt{/api/generate-therapy}: 
    It creates personalized therapy exercises based on analysis results.
    \item \texttt{/api/feedback}: 
    It provides personalized guidance and tracks progress over time.
\end{itemize}
To efficiently support this, we devise the following processing flow.  
\begin{enumerate}[leftmargin=*]
    \item The dataflow begins with the client uploading an audio file to the ``\texttt{/api/speech-analyze}'' endpoint. 
    The server processes this input through an accurate speech-to-text conversion. 
    \item The resulting text and phoneme confidence scores are analyzed by the pattern analysis module, which identifies speech issues.
    \item Then, the client requests personalized exercises for therapy by sending the analysis identifier of detected speech issues to the ``\texttt{/api/generate-therapy}'' endpoint. 
    \item Afterward, the server generates exercises tailored to the detected speech patterns and returns them to the client. 
    \item Finally, the ``\texttt{/api/feedback}'' endpoint provides guidance on improving pronunciation.
\end{enumerate}

\section{Experimental Methodology}
\label{Sec_Exp}

We evaluate the SpikeVox framework using PyTorch-based implementation, and then run it on the Apple M4 10-core CPU with 16GB memory.
SpikeVox employs the SpikeGPT model with 216M parameters~\cite{Ref_Chu_SpikeGPT_TMLR24}, that has been trained with 5B tokens from the OpenWebText dataset~\cite{pile}.  
To evaluate speech analysis performance of SpikeVox, we use an open-sourced dysfluency corpus (i.e., Libri-Dys dataset~\cite{Ref_Lian_SSDM_NeurIPS24}), and consider standard metrics, such as transcription accuracy, issue categorization, and phoneme detection.

\section{Results and Discussion}
\label{Sec_Results}

\subsection{Speech Analysis Performance}
\label{Sec_Results_Analysis}

\begin{figure}[t]
\centering
\includegraphics[width=\linewidth]{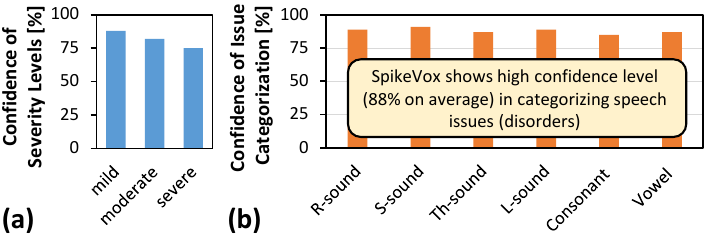}
\vspace{-0.6cm}
\caption{Performance of SpikeVox for its confidence in (a) transcription  process considering the severity level of disorders, and (b) issue categorization process.}
\label{fig_results_graph}
\vspace{-0.4cm}
\end{figure}

\textbf{Transcription.} 
Fig.~\ref{fig_results_graph}(a) shows that, our SpikeVox achieves 88\%, 82\%, and 75\%  accuracy for classifying input samples with ``mild'', ``moderate'', and ``severe'' issues. 
Such an effective classification is attributed to the high quality speech-to-text conversion using the wave2vec 2.0 library and our effective classification policy in Eq.~\ref{Eq_severity}. 
These results highlight the capability of our SpikeVox in recognizing different severity levels in the given samples. 

\textbf{Issue Categorization and Phoneme Detection.} 
One of the most critical aspect of speech therapy systems is correctly categorizing the speech issues (i.e., R-sound, S-sound, Th-sound, L-sound, consonant cluster, and vowel) based on the phoneme pronunciation, as this aspect identifies the necessary treatments to do. 
For issue categorization, our SpikeVox achieves high confidence level with 89\% for R-sound (rhotacism), 91\% for S-sound (lisping), 87\% for Th-sound, 89\% for L-sound, 85\% for consonant cluster, and 87\% for vowel (i.e., 88\% on average), as shown in Fig.~\ref{fig_results_graph}(b).
Such high performance from SpikeVox is attributed to its effective speech-to-text conversion, and effective categorization criteria presented in Table~\ref{tab_mapping_categories}. 

\subsection{SpikeVox Processing Flow}
\label{Sec_Results_Flow}

To illustrate the data processing flow in SpikeVox, we perform an experimental case study, as shown in Fig.~\ref{Fig_App}.
Label \tikzmarknode[circledColor,draw=black,fill=brown,text=black]{t1}{1} indicates the pages that SpikeVox GUI provides, namely \textit{recording}, \textit{analysis}, \textit{therapy}, and \textit{feedback}. 
In the recording page, the user can record the speech sound, which will be analyzed for its transcription and phoneme pronunciation in the analysis page. 
For instance, the input sound of ``hello good morning'' is transcribed to ``HELLO GOOD MORNING'', showing no differences in transcription \tikzmarknode[circledColor,draw=black,fill=brown,text=black]{t1}{2}.
However, when we observe the pronunciation using Eq.~\ref{Eq_confidence}, SpikeVox identifies phonemes issues for `H', `E', `L' with 70.5\%, 41.7\%, and 73.8\%, respectively; see \tikzmarknode[circledColor,draw=black,fill=brown,text=black]{t1}{3}.
This shows that, despite no transcription errors, there is a possibility for the existence of phoneme issues. 
Based on this observation, an overall assessment (including severity level and primary issues) is provided as summary, which will be used as basis of therapy exercises generation; see \tikzmarknode[circledColor,draw=black,fill=brown,text=black]{t1}{4}.
In the therapy page, SpikeVox generates multiple exercises based on previous results to train specific sounds (e.g., `L' and `vowel'), which are accompanied with the textual description and correct sound examples; see \tikzmarknode[circledColor,draw=black,fill=brown,text=black]{t1}{5}.
Results from this page will be used as basis of the feedback generation. 
In the feedback page, SpikeVox provides an assessment summary based on the user's current speech quality; see \tikzmarknode[circledColor,draw=black,fill=brown,text=black]{t1}{6}.
It also provides specific guidance based on the observed issues (e.g., `L' and `vowel'), visual pronunciation guides, and general tips for improving the quality of exercises; see \tikzmarknode[circledColor,draw=black,fill=brown,text=black]{t1}{7}.
SpikeVox also keeps the progress of the user's speech quality over time for improving the exercise quality; see \tikzmarknode[circledColor,draw=black,fill=brown,text=black]{t1}{8}.

\subsection{Computational Complexity and Energy Efficiency Benefits}
\label{Sec_Results_Benefits}

In speech therapy tools, the employment of generative NN models is important for analyzing disorders and providing proper feedback.
However, such generative models often rely on the transformer's attention mechanism which operates with quadratic computational complexity $\mathcal{O}(T^2)$.
To address this, we employ SpikeGPT which replaces the self-attention matrix multiplication with recurrent structure, that processes tokens sequentially and accumulates context through hidden states~\cite{Ref_Chu_SpikeGPT_TMLR24}, leading to lower complexity $\mathcal{O}(T)$.
Moreover, this approach also results in significant reduction of energy consumption due to: (1) reduced elementary operation energy, $\sim$5x energy saving is obtained by replacing multiplication-and-accumulation with accumulation only)~\cite{Ref_Chu_SpikeGPT_TMLR24}\cite{horowitz20141}; and (2) higher sparsity of operations, $\sim$0.15x fewer operations is obtained by employing spiking neurons~\cite{Ref_Chu_SpikeGPT_TMLR24}.

\section{Conclusion}
\label{Sec_Conclude}

In this paper, we propose a novel SpikeVox framework for enabling energy-efficient speech therapy solutions based on spike-driven generative language model. 
Its key steps include speech recognition, pattern analysis for disorder detection, exercises and guidance generation as therapy feedback. 
It also enable seamless interaction for users through REST API implementation. 
The experimental results show that SpikeVox is a promising framework that may provide accessible and energy efficient speech therapy solutions worldwide. 

\section*{Acknowledgment}
\label{Sec_Ack}

This work was partially supported by the NYUAD Center for Artificial Intelligence and Robotics (CAIR), funded by Tamkeen under the NYUAD Research Institute Award CG010.

\bibliographystyle{IEEEtran}
\bibliography{bibliography}

\begin{thebibliography}{10}
\providecommand{\url}[1]{#1}
\csname url@samestyle\endcsname
\providecommand{\newblock}{\relax}
\providecommand{\bibinfo}[2]{#2}
\providecommand{\BIBentrySTDinterwordspacing}{\spaceskip=0pt\relax}
\providecommand{\BIBentryALTinterwordstretchfactor}{4}
\providecommand{\BIBentryALTinterwordspacing}{\spaceskip=\fontdimen2\font plus
\BIBentryALTinterwordstretchfactor\fontdimen3\font minus \fontdimen4\font\relax}
\providecommand{\BIBforeignlanguage}[2]{{%
\expandafter\ifx\csname l@#1\endcsname\relax
\typeout{** WARNING: IEEEtran.bst: No hyphenation pattern has been}%
\typeout{** loaded for the language `#1'. Using the pattern for}%
\typeout{** the default language instead.}%
\else
\language=\csname l@#1\endcsname
\fi
#2}}
\providecommand{\BIBdecl}{\relax}
\BIBdecl

\bibitem{ASHA}
\BIBentryALTinterwordspacing
ASHA. Quick facts about asha. [Online]. Available: \url{https://www.asha.org/about/press-room/quick-facts/?srsltid=AfmBOoq1 6ml0ncrzPPGB7vH4fWyWpDnceIlDlSsQxjER52sNjhzR4jJD}
\BIBentrySTDinterwordspacing

\bibitem{NIDCD}
\BIBentryALTinterwordspacing
NIDCD. Quick statistics about voice, speech, language. [Online]. Available: \url{https://www.nidcd.nih.gov/health/statistics/quick-statistics-voice-speech-language}
\BIBentrySTDinterwordspacing

\bibitem{beaminghealth}
\BIBentryALTinterwordspacing
BeamingHealth. Speech therapy statistics 2025: How common are communication disorders and how well does speech therapy work? [Online]. Available: \url{https://beaminghealth.com/article/speech-therapy-statistics}
\BIBentrySTDinterwordspacing

\bibitem{AHPC}
\BIBentryALTinterwordspacing
AHPC. Registers of allied health professionals. [Online]. Available: \url{https://www.ahpc.gov.sg/for-professionals/registers-of-allied-health-professionals}
\BIBentrySTDinterwordspacing

\bibitem{Greatspeech}
\BIBentryALTinterwordspacing
GreatSpeech. What is the success rate of speech therapy? [Online]. Available: \url{https://www.greatspeech.com/what-is-the-success-rate-of-speech-therapy/}
\BIBentrySTDinterwordspacing

\bibitem{Ref_Deka_ReviewSpeechTherapy_SLH25}
C.~Deka, A.~Shrivastava, A.~K. Abraham, S.~Nautiyal, and P.~Chauhan, ``Ai-based automated speech therapy tools for persons with speech sound disorder: a systematic literature review,'' \emph{Speech, Language and Hearing}, vol.~28, no.~1, p. 2359274, 2025.

\bibitem{Ref_Bayerl_Dysfluencies_arXiv22}
S.~P. Bayerl, D.~Wagner, E.~N{\"o}th, and K.~Riedhammer, ``Detecting dysfluencies in stuttering therapy using wav2vec 2.0,'' \emph{arXiv preprint arXiv:2204.03417}, 2022.

\bibitem{Ref_Mulfari_DL4Telerehab_CMB22}
D.~Mulfari, D.~La~Placa, C.~Rovito, A.~Celesti, and M.~Villari, ``Deep learning applications in telerehabilitation speech therapy scenarios,'' \emph{Computers in Biology and Medicine (ComBioMed)}, vol. 148, p. 105864, 2022.

\bibitem{Ref_Shih_Dysarthria_Healthcare22}
D.-H. Shih, C.-H. Liao, T.-W. Wu, X.-Y. Xu, and M.-H. Shih, ``Dysarthria speech detection using convolutional neural networks with gated recurrent unit,'' in \emph{Healthcare}, vol.~10, no.~10, 2022, p. 1956.

\bibitem{Ref_Baevski_wav2vec_NeurIPS20}
A.~Baevski, Y.~Zhou, A.~Mohamed, and M.~Auli, ``wav2vec 2.0: A framework for self-supervised learning of speech representations,'' \emph{Advances in Neural Information Processing Systems (NeurIPS)}, vol.~33, pp. 12\,449--12\,460, 2020.

\bibitem{Ref_Maass_SNN_NeuNet97}
W.~Maass, ``Networks of spiking neurons: The third generation of neural network models,'' \emph{Neural Networks}, vol.~10, no.~9, pp. 1659--1671, 1997.

\bibitem{Ref_Mozafari_SpykeTorch_FNINS19}
M.~Mozafari, M.~Ganjtabesh, A.~Nowzari-Dalini, and T.~Masquelier, ``Spyketorch: Efficient simulation of convolutional spiking neural networks with at most one spike per neuron,'' \emph{Frontiers in Neuroscience}, vol.~13, p. 625, 2019.

\bibitem{Ref_Putra_FSpiNN_TCAD20}
R.~V.~W. {Putra} and M.~{Shafique}, ``Fspinn: An optimization framework for memory-efficient and energy-efficient spiking neural networks,'' \emph{IEEE Transactions on Computer-Aided Design of Integrated Circuits and Systems (TCAD)}, vol.~39, no.~11, pp. 3601--3613, 2020.

\bibitem{Ref_Putra_QSpiNN_IJCNN21}
R.~V.~W. Putra and M.~Shafique, ``Q-spinn: A framework for quantizing spiking neural networks,'' in \emph{International Joint Conference on Neural Networks (IJCNN)}.\hskip 1em plus 0.5em minus 0.4em\relax IEEE, 2021, pp. 1--8.

\bibitem{Ref_Putra_SparkXD_DAC21}
R.~V.~W. Putra, M.~A. Hanif, and M.~Shafique, ``Sparkxd: A framework for resilient and energy-efficient spiking neural network inference using approximate dram,'' in \emph{58th ACM/IEEE Design Automation Conference (DAC)}, 2021, pp. 379--384.

\bibitem{Ref_Putra_SpikeDyn_DAC21}
R.~V.~W. Putra and M.~Shafique, ``Spikedyn: A framework for energy-efficient spiking neural networks with continual and unsupervised learning capabilities in dynamic environments,'' in \emph{58th ACM/IEEE Design Automation Conference (DAC)}, 2021, pp. 1057--1062.

\bibitem{Ref_Shafique_EdgeAI_ICCAD21}
M.~Shafique, A.~Marchisio, R.~V.~W. Putra, and M.~A. Hanif, ``Towards energy-efficient and secure edge ai: A cross-layer framework iccad special session paper,'' in \emph{2021 IEEE/ACM International Conference On Computer Aided Design (ICCAD)}, 2021, pp. 1--9.

\bibitem{Ref_Putra_QSViT_arXiv25}
R.~V.~W. Putra, S.~Iftikhar, and M.~Shafique, ``Qsvit: A methodology for quantizing spiking vision transformers,'' \emph{arXiv preprint arXiv:2504.00948}, 2025.

\bibitem{Ref_Putra_SNNonCNP_arXiv25}
R.~V.~W. Putra, P.~Wickramasinghe, and M.~Shafique, ``Enabling efficient processing of spiking neural networks with on-chip learning on commodity neuromorphic processors for edge ai systems,'' \emph{arXiv preprint arXiv:2504.00957}, 2025.

\bibitem{Ref_Putra_SpikeNAS_TAI25}
R.~V.~W. Putra and M.~Shafique, ``Spikenas: A fast memory-aware neural architecture search framework for spiking neural network-based embedded ai systems,'' \emph{IEEE Transactions on Artificial Intelligence (TAI)}, pp. 1--12, 2025.

\bibitem{Ref_Akopyan_TrueNorth_TCAD15}
F.~{Akopyan}, J.~{Sawada}, A.~{Cassidy}, R.~{Alvarez-Icaza}, J.~{Arthur}, P.~{Merolla}, N.~{Imam}, Y.~{Nakamura}, P.~{Datta}, G.~{Nam}, B.~{Taba}, M.~{Beakes}, B.~{Brezzo}, J.~B. {Kuang}, R.~{Manohar}, W.~P. {Risk}, B.~{Jackson}, and D.~S. {Modha}, ``Truenorth: Design and tool flow of a 65 mw 1 million neuron programmable neurosynaptic chip,'' \emph{IEEE Transactions on Computer-Aided Design of Integrated Circuits and Systems (TCAD)}, vol.~34, no.~10, pp. 1537--1557, Oct 2015.

\bibitem{Ref_Davies_Loihi_MM18}
M.~{Davies}, N.~{Srinivasa}, T.~{Lin}, G.~{Chinya}, Y.~{Cao}, S.~H. {Choday}, G.~{Dimou}, P.~{Joshi}, N.~{Imam}, S.~{Jain}, Y.~{Liao}, C.~{Lin}, A.~{Lines}, R.~{Liu}, D.~{Mathaikutty}, S.~{McCoy}, A.~{Paul}, J.~{Tse}, G.~{Venkataramanan}, Y.~{Weng}, A.~{Wild}, Y.~{Yang}, and H.~{Wang}, ``Loihi: A neuromorphic manycore processor with on-chip learning,'' \emph{IEEE Micro}, vol.~38, no.~1, pp. 82--99, Jan 2018.

\bibitem{Ref_Frenkel_ODIN_TBCAS19}
C.~{Frenkel}, M.~{Lefebvre}, J.~{Legat}, and D.~{Bol}, ``A 0.086-mm$^2$ 12.7-pj/sop 64k-synapse 256-neuron online-learning digital spiking neuromorphic processor in 28-nm cmos,'' \emph{IEEE TBCAS}, vol.~13, no.~1, pp. 145--158, Feb 2019.

\bibitem{Ref_SynSense_DYNAP}
\BIBentryALTinterwordspacing
SynSense. Dynap-cnn: The world’s first fully scalable, event-driven neuromorphic processor with up to 1m configurable spiking neurons and direct interface with external dvs. [Online]. Available: \url{https://www.synsense.ai/products/dynap-cnn/}
\BIBentrySTDinterwordspacing

\bibitem{Ref_BrainChip_Akida}
\BIBentryALTinterwordspacing
BrainChip. Akida neural processor soc. [Online]. Available: \url{https://brainchip.com/akida-neural-processor-soc/}
\BIBentrySTDinterwordspacing

\bibitem{Ref_Chu_SpikeGPT_TMLR24}
\BIBentryALTinterwordspacing
R.-J. Zhu, Q.~Zhao, G.~Li, and J.~Eshraghian, ``Spike{GPT}: Generative pre-trained language model with spiking neural networks,'' \emph{Transactions on Machine Learning Research (TNLR)}, 2024. [Online]. Available: \url{https://openreview.net/forum?id=gcf1anBL9e}
\BIBentrySTDinterwordspacing

\bibitem{Ref_Gao_Pile_arXiv20}
L.~Gao, S.~Biderman, S.~Black, L.~Golding, T.~Hoppe, C.~Foster, J.~Phang, H.~He, A.~Thite, N.~Nabeshima \emph{et~al.}, ``The pile: An 800gb dataset of diverse text for language modeling,'' \emph{arXiv preprint arXiv:2101.00027}, 2020.

\bibitem{Ref_Izhikevich_CompareModels_TNN04}
E.~M. {Izhikevich}, ``Which model to use for cortical spiking neurons?'' \emph{IEEE Transactions on Neural Networks (TNN)}, vol.~15, no.~5, pp. 1063--1070, Sep. 2004.

\bibitem{Ref_Putra_ReSpawn_ICCAD21}
R.~V.~W. Putra, M.~A. Hanif, and M.~Shafique, ``Respawn: Energy-efficient fault-tolerance for spiking neural networks considering unreliable memories,'' in \emph{IEEE/ACM International Conference On Computer Aided Design (ICCAD)}, 2021, pp. 1--9.

\bibitem{Ref_Putra_lpSpikeCon_IJCNN22}
R.~V.~W. Putra and M.~Shafique, ``lpspikecon: Enabling low-precision spiking neural network processing for efficient unsupervised continual learning on autonomous agents,'' in \emph{International Joint Conference on Neural Networks (IJCNN)}, 2022, pp. 1--8.

\bibitem{Ref_Putra_SoftSNN_DAC22}
R.~V.~W. Putra, M.~A. Hanif, and M.~Shafique, ``Softsnn: Low-cost fault tolerance for spiking neural network accelerators under soft errors,'' in \emph{59th ACM/IEEE Design Automation Conference (DAC)}, 2022, pp. 151--156.

\bibitem{Ref_Putra_EnforceSNN_FNINS22}
------, ``Enforcesnn: Enabling resilient and energy-efficient spiking neural network inference considering approximate drams for embedded systems,'' \emph{Frontiers in Neuroscience (FNINS)}, vol.~16, p. 937782, 2022.

\bibitem{Ref_Putra_TopSpark_IROS23}
R.~V.~W. Putra and M.~Shafique, ``Topspark: a timestep optimization methodology for energy-efficient spiking neural networks on autonomous mobile agents,'' in \emph{IEEE/RSJ International Conference on Intelligent Robots and Systems (IROS)}.\hskip 1em plus 0.5em minus 0.4em\relax IEEE, 2023, pp. 3561--3567.

\bibitem{Ref_Rathi_SNNsurvey_CSUR23}
N.~Rathi, I.~Chakraborty, A.~Kosta, A.~Sengupta, A.~Ankit, P.~Panda, and K.~Roy, ``Exploring neuromorphic computing based on spiking neural networks: Algorithms to hardware,'' \emph{ACM Compututing Survey}, vol.~55, no.~12, 2023.

\bibitem{Ref_Diehl_STDPmnist_FNCOM15}
P.~Diehl and M.~Cook, ``Unsupervised learning of digit recognition using spike-timing-dependent plasticity,'' \emph{Frontiers in Computational Neuroscience}, vol.~9, p.~99, 2015.

\bibitem{Ref_Neftci_SurrogateSNNs_IEEEMSP19}
E.~O. Neftci, H.~Mostafa, and F.~Zenke, ``Surrogate gradient learning in spiking neural networks: Bringing the power of gradient-based optimization to spiking neural networks,'' \emph{IEEE Signal Processing Magazine}, vol.~36, no.~6, pp. 51--63, 2019.

\bibitem{Ref_Shriberg_DisorderClass_CLP10}
L.~D. Shriberg, M.~Fourakis, S.~D. Hall, H.~B. Karlsson, H.~L. Lohmeier, J.~L. McSweeny, N.~L. Potter, A.~R. Scheer-Cohen, E.~A. Strand, C.~M. Tilkens \emph{et~al.}, ``Extensions to the speech disorders classification system (sdcs),'' \emph{Clinical Linguistics \& Phonetics}, vol.~24, no.~10, pp. 795--824, 2010.

\bibitem{pile}
L.~Gao, S.~Biderman, S.~Black, L.~Golding, T.~Hoppe, C.~Foster, J.~Phang, H.~He, A.~Thite, N.~Nabeshima, S.~Presser, and C.~Leahy, ``The {P}ile: An 800gb dataset of diverse text for language modeling,'' \emph{arXiv preprint arXiv:2101.00027}, 2020.

\bibitem{Ref_Lian_SSDM_NeurIPS24}
J.~Lian, X.~Zhou, Z.~Ezzes, J.~Vonk, B.~Morin, D.~P. Baquirin, Z.~Miller, M.~L. Gorno~Tempini, and G.~Anumanchipalli, ``Ssdm: Scalable speech dysfluency modeling,'' \emph{Advances in Neural Information Processing Systems (NeurIPS)}, vol.~37, pp. 101\,818--101\,855, 2024.

\bibitem{horowitz20141}
M.~Horowitz, ``1.1 computing's energy problem (and what we can do about it),'' in \emph{2014 IEEE International Solid-State Circuits Conference Digest of Technical Papers (ISSCC)}.\hskip 1em plus 0.5em minus 0.4em\relax IEEE, 2014, pp. 10--14.

\end{thebibliography}
\end{spacing}

\end{document}